\documentclass[10pt,conference]{IEEEtran}
\usepackage[T1]{fontenc}
\usepackage{lmodern}
\usepackage[utf8]{inputenc}
\usepackage{url}
\usepackage{breakurl} 
\usepackage[bookmarks=true, bookmarksnumbered=true, hypertexnames=false, breaklinks=true, linkbordercolor={1 1 1}, pdfborder={0 0 0}]{hyperref}

\let\labelindent\relax
\usepackage{enumitem}
\usepackage{microtype}
\newcommand{\rquestion}{What are the experienced consequences of unhappiness among software developers while developing software?}

\begin{document}
%
\title{Unhappy Developers: Bad for Themselves, Bad for Process, and Bad for Software Product}

\author{
    \IEEEauthorblockN{
        Daniel Graziotin\IEEEauthorrefmark{1},
        Fabian Fagerholm\IEEEauthorrefmark{2}, 
        Xiaofeng Wang\IEEEauthorrefmark{3} and
        Pekka Abrahamsson\IEEEauthorrefmark{4}
    }
    \IEEEauthorblockA{
    	\IEEEauthorrefmark{1}Institute of Software Technology, University of Stuttgart, Germany, \url{daniel.graziotin@informatik.uni-stuttgart.de}
    }
    \IEEEauthorblockA{
    	\IEEEauthorrefmark{2}Department of Computer Science, University of Helsinki, Finland, \url{fabian.fagerholm@helsinki.fi}
    }
    \IEEEauthorblockA{
    	\IEEEauthorrefmark{3}Faculty of Computer Science, Free University of Bozen-Bolzano, Italy, \url{xiaofeng.wang@unibz.it}
    }
    \IEEEauthorblockA{
    	\IEEEauthorrefmark{4}Department of Computer and Information Science (IDI), NTNU, Trondheim, Norway, \url{pekkaa@ntnu.no}
    }
}

\maketitle

\begin{abstract}
Recent research in software engineering supports the ``happy-productive'' thesis, and the desire of flourishing happiness among programmers is often expressed by industry practitioners. Recent literature has suggested that a cost-effective way to foster happiness and productivity among workers could be to limit unhappiness of developers due to its negative impact. However, possible negative effects of unhappiness are still largely unknown in the software development context. In this paper, we present the first results from a study exploring the consequences of the unhappy developers. Using qualitative data analysis of the survey responses given by 181 participants, we identified 49 potential consequences of unhappiness while developing software. These results have several implications. While raising the awareness of the role of moods, emotions and feelings in software development, we foresee that our classification scheme will spawn new happiness studies linking causes and effects, and it can act as a guideline for developers and managers to foster happiness at work.
\end{abstract}


%
\IEEEpeerreviewmaketitle

\section{Introduction}
A practice that has emerged recently is to promote flourishing happiness among workers in order to enact the happy-productive worker thesis \cite{Zelenski2008}. Relevant research in software engineering (SE) has recently emerged which considers the positive aspects of happiness and affect in general, e.g.,\cite{graziotin2015you,Graziotin2014PEERJ,graziotin2015feelings,Fagerholm2015,Muller2015,Ortu2015a,Novielli:2015km,Khan2010,Destefanis2016}. Yet, software developers are prone to share horror stories about their working experience on a daily basis \cite{Graziotin2014IEEESW}.

As objective happiness is suggested to be bounded by the difference between experienced positive and negative affect \cite{Diener:1999cl,Kahneman:1999ck}, one way for maximizing happiness is to understand and minimize the negative experiences. There are calls to understand the benefits of limiting negative experiences on the job \cite{Diener:1999cl}, and SE research has yet to explore the research avenue. Based on the existing literature, we set the following research question: \\
RQ:  \emph{\rquestion}

We are conducting a series of studies in the context of a large-scale quantitative and qualitative survey of software developers. In this paper, we present our analysis of 181 complete responses related to the consequences of unhappiness while developing software. We report an openly available schema \cite{Graziotin:2016ck} of 49 consequences of unhappiness that we identified.

\section{Method}
\label{sec:method:method}

The study we are presenting is the first of a series of inquiries that we conducted through a large-scale survey of GitHub developers. The overall research project employs a mixed research method, comprising both elements of quantitative and qualitative research~\cite{Creswell2009}; in this paper, we address the present RQ using the qualitative data collected through the survey.

We retrieved the contact information of GitHub contributors using the publicly available data at the GitHub Archive~\cite{Grigorik:2016}, which collects and stores a log of public events occurring in GitHub. The questionnaire used in the overall survey was composed of (1) questions to collect demographic information, (2) one question carrying the Scale of Positive and Negative Experience (SPANE~\cite{Diener2010, Diener2009a}, which we described in~\cite{Graziotin2015SSE}), 12 scale items for assessing the happiness of an individual, and (3) two open-ended questions on the causes and consequences of positive and negative affect while developing software. We describe the questionnaire systematically in an online appendix~\cite{Graziotin:2016ck}.

In order to answer the RQ, we developed a coding strategy to analyze the responses to the open-ended question related to negative affect. We applied open coding, axial coding, and selective coding \cite{Corbin2008}. A working example of the coding process is available online \cite{Graziotin:2016ck}.

\section{Results}
\label{sec:results:results}

After data cleaning, we obtained $181$ valid and complete responses related to our RQ. Among the $181$ participants, 78\% were professional software developers, 13\% were in other roles (e.g., CEO, CTO, academic researcher), and the remaining were students or unemployed. The average year of birth was 1984 (standard deviation (sd)=8.27, median 1986).  The average experience with software development was 8.22 years (sd=7.83, median=5).

We identified 254 codes related to the consequences of unhappiness, which resulted in 49 types of consequences, further grouped into 16 categories and sub-categories. Because of the space limitation, we report here the most frequent codes only. The entire dataset is available as archived open data \cite{Graziotin:2016ck}.

\subsubsection{Internal Consequences---Developer's Own Being}
\label{sssec:resultsrq3:individual_consequences}

\textbf{Low cognitive performance} refers to low mental performance, such as low focus: ``\textit{[\ldots] the negative feelings lead to not thinking things through as clearly as I would have if the feeling of frustration was not present}''; cognitive skills dropping off: ``\textit{My software dev skills dropped off as I became more and more frustrated until I eventually closed it off and came back the next day to work on it}''; and general mental fatigue: ``\textit{Getting frustrated and sloppy}''.

The \textbf{mental unease or disorder} category collects consequences related to mental health. Unhappiness while developing software is a cause of, in decreasing order of frequency, stress and burnout: ``\textit{[\ldots] only reason of my failure due of burnout}''; anxiety: ``\textit{These kinds of situations make me feel panicky}''; low self-esteem: ``\textit{If I feel particularly lost on a certain task, I may sometimes begin to question my overall ability to be a good programmer}''; and sadness, if not depression: ``\textit{feels like a black fog of depression surrounds you and the project}''.

\textbf{Low motivation}. The participants were clear in stating that unhappiness led to low motivation for developing software, e.g, ``\textit{[the unhappiness] has left me feeling very stupid and as a result I have no leadership skills, no desire to participate and feel like I'm being forced to code to live as a kind of punishment. [\ldots]}'', or ``\textit{Also, I'm working at a really slow pace [\ldots] because I'm just not as engaged with the work}''.

\subsubsection{External Consequences}

\paragraph{Process Consequences}

The \textit{process} category collects those consequences that are related to a software development process, endeavor, or set of practices that are not explicitly tied up to an artifact.

\textbf{Low productivity} groups all consequences of unhappiness related to performance and productivity losses, as we defined previously \cite{graziotin2015you,Fagerholm2015}. Some participants made very simple and clear statements such as ``\textit{productivity drops}'' or ``\textit{[negative experience] definitely makes me work slower}''. Others elaborated more: ``\textit{[unhappiness] made it harder or impossible to come up with solutions}'', ``\textit{[\ldots], and [the negative experience] slowed my progress because of the negative feeling toward the feature}''.

Unhappiness was reported to cause \textbf{delay} in process execution: ``\textit{In both cases [negative experiences] the emotional toll on me caused delays to the project}''. It also causes glitches in communication and a disorganized process.

Developers declared that unhappiness caused them to \textbf{deviate from the process} or the agreed set of practices. Specifically, unhappiness makes developers compromise in terms of actions, in order to just get rid of tasks: ``\textit{In these instances my development tended towards immediate and quick 'ugly' solutions}''. They decide to take shortcuts when enacting a software process, compromising the quality of the process itself.

\paragraph{Artifact-oriented Consequences}
\label{sssec:resultsrq3:artifactoriented_consequences}
The category of \textit{artifact-oriented} consequences groups all those consequences that are directly related to a development product, e.g., software code, requirements, and to working with it.

\textbf{Low code quality} represents the consequences of unhappiness of developers that are related to deterioration of the artifacts' quality. The participants reported that ``\textit{eventually [due to negative experiences], code quality cannot be assured. So this will make my code messy and more bug can be found in it}'', and ``\textit{As a result my code becomes sloppier}''.

\textbf{Discharging code} could be seen as an extreme case of productivity and quality deterioration. We found in some instances that the participants destroyed the task-related codebase, e.g, ``\textit{I deleted the code that I was writing because I was a bit angry}'', up to deleting the entire projects: ``\textit{I have deleted entire projects to start over with code that didn't seem to be going in a wrong direction}''.

\section{Discussion and Conclusion}\label{sec:discussion:discussion}

We presented the results of an analysis of the experienced consequences of unhappiness among software developers at work.

We found that the highest impact of unhappiness is on productivity and performance, which, by aggregating the categories of low cognitive performance and process-related productivity, accounts for about 40\% of the in-text references. This is in line with all the studies that concentrate on the relationship between affect and performance in SE \cite{graziotin2015you, Graziotin2014PEERJ, graziotin2015feelings, Muller2015, Khan2010, Wrobel2013} and quantify the relationship. 

We also found that unhappiness causes issues related to mental health. Those issues include situations of mental unease, such as low self-esteem, high anxiety, burnout, and stress, but also possible disorders such as depression. Unhappiness also leads to low motivation among developers, which is a critical issue in SE \cite{Franca2014b}, as well as work withdrawal, even to the extent of quitting jobs. Investigation of these relationships is ongoing in psychology research \cite{Miner2010} but are currently not adressed in SE research.

Finally, our results show that unhappiness makes developers take shortcuts in the process, often leading to low code quality. The results enforce several initial studies \cite{Khan2010,Destefanis2016,Ortu:2016gz,Ortu2015a} that attempt to understand the relationship between developers' affect and software quality.

Our research has found a plethora of consequences of unhappiness that are of interest to practitioners regardless of their roles. We summarized the most prominent ones in the present paper, but practitioners could be interested in the complete list that is available as open data \cite{Graziotin:2016ck}. 

Intervening on the affect of developers might have relatively low cost but astonishing benefits \cite{graziotin2015you}. Our study has expanded the previously gathered knowledge by suggesting that limiting unhappiness will limit the damage in terms of several factors from individual, artifact, and process perspectives.

\IEEEtriggeratref{13}


%
\bibliographystyle{IEEEtran}
\bibliography{references}

\begin{thebibliography}{10}
\providecommand{\url}[1]{#1}
\csname url@samestyle\endcsname
\providecommand{\newblock}{\relax}
\providecommand{\bibinfo}[2]{#2}
\providecommand{\BIBentrySTDinterwordspacing}{\spaceskip=0pt\relax}
\providecommand{\BIBentryALTinterwordstretchfactor}{4}
\providecommand{\BIBentryALTinterwordspacing}{\spaceskip=\fontdimen2\font plus
\BIBentryALTinterwordstretchfactor\fontdimen3\font minus
  \fontdimen4\font\relax}
\providecommand{\BIBforeignlanguage}[2]{{%
\expandafter\ifx\csname l@#1\endcsname\relax
\typeout{** WARNING: IEEEtran.bst: No hyphenation pattern has been}%
\typeout{** loaded for the language `#1'. Using the pattern for}%
\typeout{** the default language instead.}%
\else
\language=\csname l@#1\endcsname
\fi
#2}}
\providecommand{\BIBdecl}{\relax}
\BIBdecl

\bibitem{Zelenski2008}
J.~M. Zelenski, S.~A. Murphy, and D.~A. Jenkins, ``{The Happy-Productive Worker
  Thesis Revisited},'' \emph{Journal of Happiness Studies}, vol.~9, no.~4, pp.
  521--537, 2008.

\bibitem{graziotin2015you}
D.~Graziotin, X.~Wang, and P.~Abrahamsson, ``{How do you feel, developer? An
  explanatory theory of the impact of affects on programming performance},''
  \emph{PeerJ Computer Science}, vol.~1, no.~1, p. e18, Aug. 2015.

\bibitem{Graziotin2014PEERJ}
------, ``{Happy software developers solve problems better: psychological
  measurements in empirical software engineering},'' \emph{PeerJ}, vol.~2, p.
  e289, 2014.

\bibitem{graziotin2015feelings}
------, ``{Do feelings matter? On the correlation of affects and the
  self-assessed productivity in software engineering},'' \emph{Journal of
  Software: Evolution and Process}, vol.~27, no.~7, pp. 467--487, 2014.

\bibitem{Fagerholm2015}
F.~Fagerholm, M.~Ikonen, P.~Kettunen, J.~M{\"u}nch, V.~Roto, and
  P.~Abrahamsson, ``{Performance Alignment Work: How software developers
  experience the continuous adaptation of team performance in Lean and Agile
  environments},'' \emph{Information and Software Technology}, vol.~64, pp.
  132--147, 2015.

\bibitem{Muller2015}
S.~C. Muller and T.~Fritz, ``{Stuck and Frustrated or in Flow and Happy:
  Sensing Developers' Emotions and Progress},'' in \emph{2015 IEEE/ACM 37th
  IEEE International Conference on Software Engineering}.\hskip 1em plus 0.5em
  minus 0.4em\relax IEEE, May 2015, pp. 688--699.

\bibitem{Ortu2015a}
M.~Ortu, B.~Adams, G.~Destefanis, P.~Tourani, M.~Marchesi, and R.~Tonelli,
  ``{Are Bullies More Productive? Empirical Study of Affectiveness vs. Issue
  Fixing Time},'' in \emph{2015 IEEE/ACM 12th Working Conference on Mining
  Software Repositories (MSR)}.\hskip 1em plus 0.5em minus 0.4em\relax IEEE,
  2015, pp. 303--313.

\bibitem{Novielli:2015km}
N.~Novielli, F.~Calefato, and F.~Lanubile, ``{The challenges of sentiment
  detection in the social programmer ecosystem},'' in \emph{SSE 2015
  Proceedings of the 7th International Workshop on Social Software
  Engineering}.\hskip 1em plus 0.5em minus 0.4em\relax New York, New York, USA:
  ACM, Sep. 2015.

\bibitem{Khan2010}
I.~A. Khan, W.-P. Brinkman, and R.~M. Hierons, ``{Do moods affect
  programmers{\textquoteright} debug performance?}'' \emph{Cognition,
  Technology {\&} Work}, vol.~13, no.~4, pp. 245--258, 2010.

\bibitem{Destefanis2016}
G.~Destefanis, M.~Ortu, S.~Counsell, S.~Swift, M.~Marchesi, and R.~Tonelli,
  ``{Software development: do good manners matter?}'' \emph{PeerJ Computer
  Science}, vol.~2, no.~4, p. e73, 2016.

\bibitem{Graziotin2014IEEESW}
D.~Graziotin, X.~Wang, and P.~Abrahamsson, ``{Software Developers, Moods,
  Emotions, and Performance.}'' \emph{IEEE Software}, vol.~31, no.~4, pp.
  24--27, 2014.

\bibitem{Diener:1999cl}
E.~Diener, E.~M. Suh, R.~E. Lucas, and H.~L. Smith, ``{Subjective well-being:
  Three decades of progress.}'' \emph{Psychological Bulletin}, vol. 125, no.~2,
  pp. 276--302, 1999.

\bibitem{Kahneman:1999ck}
D.~Kahneman, ``{Objective Happiness},'' in \emph{Well-Being: Foundations of
  Hedonic Psychology}, D.~Kahneman, E.~Diener, and N.~Schwarz, Eds.\hskip 1em
  plus 0.5em minus 0.4em\relax New York, NY, USA: Utilitas, 1999, pp. 3--25.

\bibitem{Graziotin:2016ck}
D.~Graziotin, F.~Fagerholm, X.~Wang, and P.~Abrahamsson, ``{Online appendix:
  the happiness of software developers},'' \emph{figshare}, 2017,
  \url{https://doi.org/10.6084/m9.figshare.c.3355707}.

\bibitem{Creswell2009}
J.~W. Creswell, \emph{{Research design: qualitative, quantitative, and mixed
  method approaches}}, 3rd~ed.\hskip 1em plus 0.5em minus 0.4em\relax Thousand
  Oaks, California: Sage Publications, 2009, vol. 2nd.

\bibitem{Grigorik:2016}
\BIBentryALTinterwordspacing
I.~Grigorik. (2012) {GitHub Archive}. [Online]. Available:
  \url{https://githubarchive.org}
\BIBentrySTDinterwordspacing

\bibitem{Diener2010}
E.~Diener, D.~Wirtz, W.~Tov, C.~Kim-Prieto, D.-w. Choi, S.~Oishi, and
  R.~Biswas-Diener, ``{New Well-being Measures: Short Scales to Assess
  Flourishing and Positive and Negative Feelings},'' \emph{Social Indicators
  Research}, vol.~97, no.~2, pp. 143--156, May 2010.

\bibitem{Diener2009a}
\BIBentryALTinterwordspacing
E.~Diener, D.~Wirtz, W.~Tov, C.~Kim-Prieto, D.~Choi, S.~Oishi, and
  R.~Biswas-Diener. (2009) {Scale of Positive and Negative Experience (SPANE)}.
  [Online]. Available:
  \url{http://internal.psychology.illinois.edu/~ediener/SPANE.html}
\BIBentrySTDinterwordspacing

\bibitem{Graziotin2015SSE}
D.~Graziotin, X.~Wang, and P.~Abrahamsson, ``{Understanding the affect of
  developers: theoretical background and guidelines for psychoempirical
  software engineering},'' in \emph{SSE 2015: Proceedings of the 7th
  International Workshop on Social Software Engineering}.\hskip 1em plus 0.5em
  minus 0.4em\relax ACM, Sep. 2015, pp. 25--32.

\bibitem{Corbin2008}
J.~M. Corbin and A.~L. Strauss, \emph{{Basics of Qualitative Research:
  Techniques and Procedures for Developing Grounded Theory}}, 3rd~ed., ser.
  Basics of Qualitative Research: Techniques and Procedures for Developing
  Grounded Theory.\hskip 1em plus 0.5em minus 0.4em\relax London: Sage
  Publications, 2008, vol. 2nd.

\bibitem{Wrobel2013}
M.~R. Wrobel, ``{Emotions in the software development process},'' \emph{2013
  6th International Conference on Human System Interactions (HSI)}, pp.
  518--523, 2013.

\bibitem{Franca2014b}
C.~Fran{\c c}a, H.~Sharp, and F.~da~Silva, ``{Motivated software engineers are
  engaged and focused, while satisfied ones are happy},'' in \emph{Proceedings
  of the 8th ACM/IEEE International Symposium on Empirical Software Engineering
  and Measurement}.\hskip 1em plus 0.5em minus 0.4em\relax New York, New York,
  USA: ACM Press, 2014, pp. 1--8.

\bibitem{Miner2010}
A.~G. Miner and T.~M. Glomb, ``{State mood, task performance, and behavior at
  work: A within-persons approach},'' \emph{Organizational Behavior and Human
  Decision Processes}, vol. 112, no.~1, pp. 43--57, May 2010.

\bibitem{Ortu:2016gz}
M.~Ortu, G.~Destefanis, S.~Counsell, S.~Swift, M.~Marchesi, and R.~Tonelli,
  ``{How diverse is your team? Investigating gender and nationality diversity
  in GitHub teams},'' \emph{Peerj Preprints}, no.~4, p. e2285v1, Jul. 2016.

\end{thebibliography}

\end{document}